\documentclass[]{pasj02} 

\usepackage[switch,mathlines]{lineno} 
\usepackage{url}
\usepackage{graphicx}
\let\degree\relax
\usepackage{gensymb}
\usepackage{color}

\jyear{2025}
\Received{}
\Accepted{}

\begin{document} 
\nolinenumbers

\title{The Subaru-Asahi StarCam: Description of the System}

\author{Ichi
\textsc{Tanaka},\altaffilmark{1}\altemailmark\orcid{0000-0002-4937-4738} \email{ichi@naoj.org} 
 Masanobu \textsc{Higashiyama},\altaffilmark{2}
 Masayo \textsc{Nakajima},\altaffilmark{1}
Toyokazu \textsc{Uda},\altaffilmark{3}
 Hitoshi \textsc{Hasegawa},\altaffilmark{4}
 Mikiya \textsc{Sato},\altaffilmark{5}
 and 
 Jun-ichi \textsc{Watanabe}\altaffilmark{5}
}
\altaffiltext{1}{Subaru Telescope, National Astronomical Observatory of Japan, 650 North A'ohoku Place, Hilo, Hawaii, 96720, U.S.A.}
\altaffiltext{2}{The Asahi Shimbun Company, 5-3-2 Tsukiji, Chuo-ku, Tokyo, Japan}
\altaffiltext{3}{Aiharasoft, Co., Sagamihara, Kanagawa 252-0141, Japan}
\altaffiltext{4}{Suginami-ku, Tokyo 168-0074, JAPAN}
\altaffiltext{5}{National Astronomical Observatory of Japan, 2-21-1 Osawa, Mitaka, Tokyo 181-8588, Japan}



\maketitle

\begin{abstract}
The Subaru-Asahi StarCam is a high-sensitivity live-streaming camera for meteor observation, installed on the dome of the Subaru Telescope at the summit area of Maunakea, Hawai'i. Although it was originally intended to share the Maunakea night sky with the public, including the local Hawai'i community, the system quickly demonstrated its potential for scientific research, owing to its highly sensitive video capabilities and the exceptional fraction of clear nights at the site. The core of the StarCam system features a Sony FX3 camera body paired with an F1.4 wide-angle lens, offering a field of view of 70$^\circ \times$ 40$^\circ$. Leveraging a state-of-the-art, high-sensitivity CMOS sensor and a bright lens, the system is capable of capturing stars as faint as magnitude 8 in real-time, with an effective frame rate of 15–-30 fps. Live streaming via YouTube began in April 2021, and the feed is constantly monitored by more than a hundred viewers at any given nighttime. This has enabled the camera to be used not only for observing regular meteor showers but also for monitoring scientifically important phenomena such as fireballs or unexpected meteor outbursts. Notable scientific achievements include: 1) Detection of the new Arid meteor shower in 2021, 2) Identification of a sub-peak activity in the $\gamma$-Perseid meteor shower (2021), 3) Detection of the 2022 $\tau$-Herculid meteor shower outburst, 4) Confirmation of the activity of the Andromedid meteor shower (2021), and 5) Multiple detections of meteor cluster phenomena. We discuss the potential and the future scope of StarCam as an open-access, real-time data platform for citizen science in meteor observations.
\end{abstract}


\section{Introduction}
The astronomical phenomena captured serendipitously by citizen cameras are increasingly contributing to scientific research. For instance, videos of fireballs recorded by the general public have enabled the estimation of their trajectories and, in some cases, led to the discovery of meteorites (e.g., \cite{TR2010}; \cite{2012Sci...338.1583J}; \cite{2013arXiv1303.1796Z}; \cite{2021M&PS...56..425B}). Furthermore, the incidence of fortuitous events, such as the atmospheric reentry of rockets and artificial satellites, being captured by cameras and reported through social media platforms has risen. This trend can be attributed to substantial advancements in semiconductor detector technology, which have significantly enhanced low-light sensitivity, along with the widespread adoption of devices like smartphones and dashboard cameras equipped with high-performance sensors. Not only have these fortuitous events been captured, but also the routine observation and recording of the night sky has become increasingly popular. This development holds revolutionary implications for the field of astronomical education and outreach, as it enables individuals residing in areas with significant light pollution or those constrained by time zones to readily observe dark skies and rare astronomical phenomena via the internet. Effective utilization of these resources can broaden access to astronomical experiences and serve as a catalyst for fostering public interest in astronomy (e.g., \cite{2017AstRv..13...28G}). 

In the field of meteor science, several dedicated video observation networks have been established, comprising multiple interconnected meteor observation systems (\cite{2009JIMO...37...55S}; \cite{2011Icar..216...40J}; \cite{2013Icar..225..614W}; \cite{2021MNRAS.506.5046V}). However, the cameras employed in these networks are not always state-of-the-art, high-sensitivity systems. The integration of rapidly evolving sensor technology into such multi-station observation systems holds the potential to catalyze novel research in meteor observation.

The 'Subaru-Asahi StarCam' (hereafter the StarCam\footnote{\url{https://www.naoj.org/PIO/LiveCam/cam\_redirect.html}}), installed on the Catwalk of the Subaru Telescope dome at Maunakea, was originally deployed for the purposes of astronomy outreach and astronomical observation environment monitoring\footnote{This is a collaborative Outreach Project between National Astronomical Observatory of Japan (NAOJ) and the Asahi Shimbun Newspaper Company. }. However, thanks to the world-class observing conditions, high percentage of clear skies, and a state-of-the-art, high-sensitivity camera, it has continued to contribute significantly to meteor science. This paper discusses the hardware overview, with several science highlights to be discussed in subsequent papers in this special issue.

\section{Background and the Operation Condition of the StarCam }\label{sec:2}
We begin by describing the background of the StarCam installation site. The harsh natural environment of Maunakea sometimes imposes limitations on hardware specifications. Maunakea is a culturally significant site for Native Hawaiians, and is therefore strictly managed by the Center for Maunakea Stewardship. We went through the necessary permission procedure upon installation of the camera. 

\subsection{Background Information of the StarCam}\label{ssec:2_1}
As mentioned in the Introduction, the primary purpose of the deployment was for site monitoring and public outreach. Maunakea is one of the world's premier observing sites, and the utility of a high-sensitivity camera for monitoring its operational environment is self-evident. The second objective pertains to the cultural context surrounding Maunakea. The mountain holds sacred significance for Native Hawaiians, and since the beginning of the 21st century, in particular, astronomical activities have been a persistent source of friction with the Native Hawaiians community. Despite efforts by astronomers to emphasize Maunakea's excellence for astronomical observations, many residents of Hawai'i remain unfamiliar with its value due to a lack of direct experience.

Maunakea rises over 4000 meters in elevation, and access to its summit area is restricted to individuals 13 years of age and older by the Center for Maunakea Stewardship. Furthermore, as an international astronomical observatory site, the summit is closed to the public-—including local residents-—after sunset\footnote{\url{https://hilo.hawaii.edu/maunakea/visitor-information/public-safety#rules}}. As a result, local residents cannot experience the night sky at Maunakea, a world-class observing site, which has essentially become an area used only by astronomers. We believe that sharing live images of Maunakea’s night sky and natural beauty with the local community can help bridge this gap and foster mutual understanding.

\subsection{Overview of the Installation Site}\label{ssec:2_2}
\begin{figure}
\centering
\includegraphics[width=0.8\linewidth]{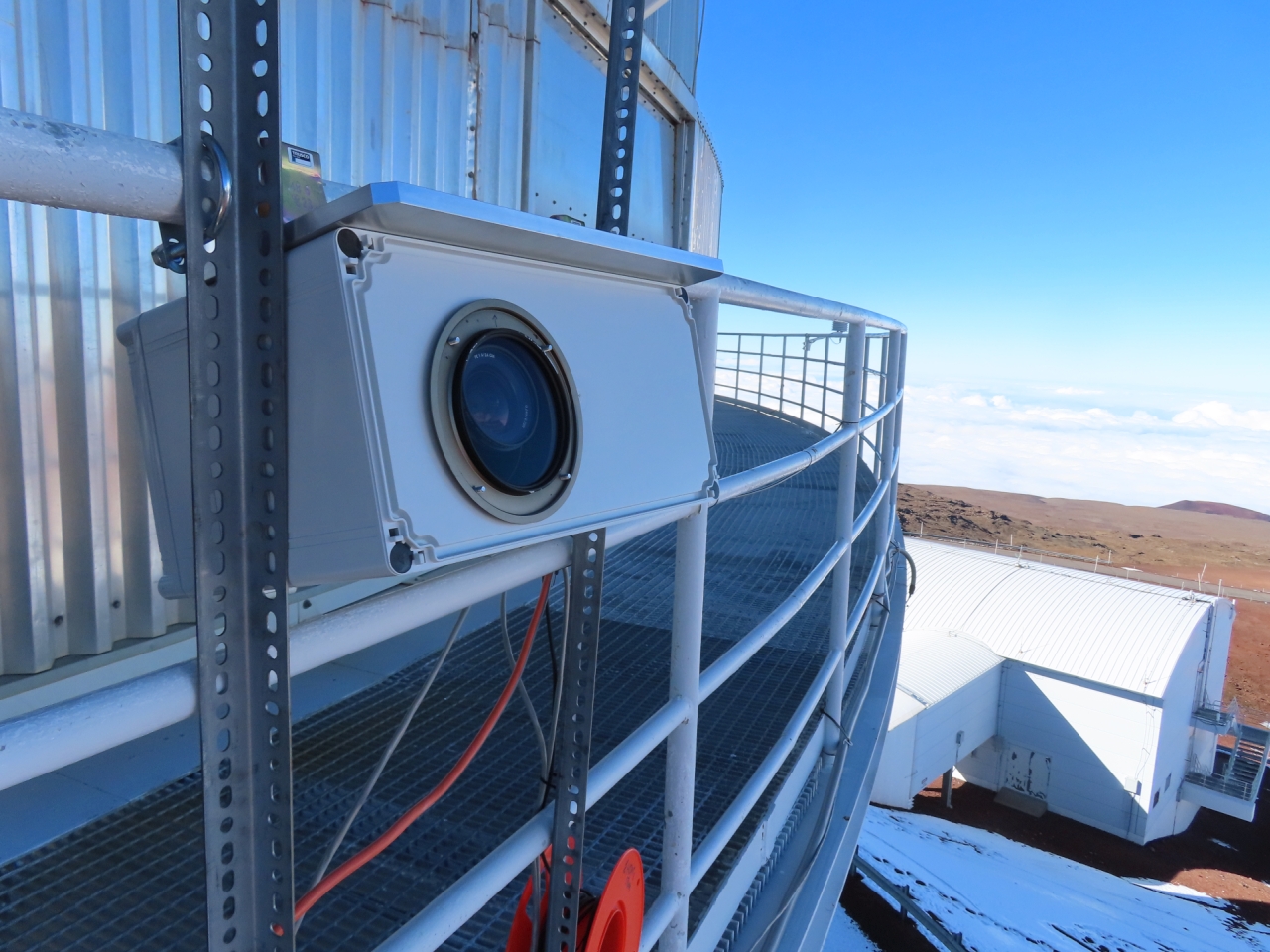}
\caption{Exterior view of the Subaru-Asahi StarCam, installed on the catwalk of the Subaru Telescope dome.
{Alt text: The exterior view of Subaru-Asahi StarCam perched on the catwalk of the Subaru telescope dome. } 
}\label{fig:fig1}
\end{figure}

\begin{figure}
\centering
\includegraphics[width=0.95\linewidth]{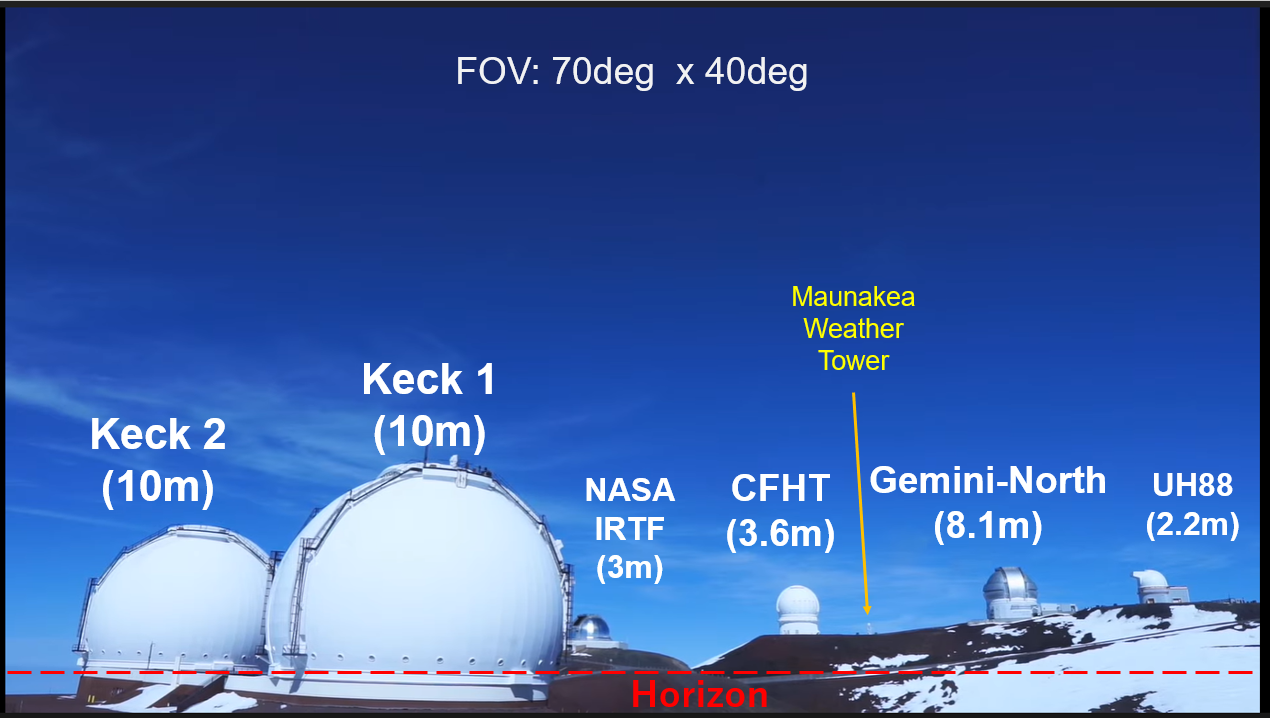}
\caption{Field of view from the Subaru-Asahi StarCam. Labels indicate visible observatory facilities; numbers in parentheses denote the size of the primary mirror (in meters).
{Alt text: This figure describes the structures visible in the camera view. } 
}\label{fig:fig2}
\end{figure}

This section summarizes the camera's installation location and operational environment. The camera is mounted on the handrail of the catwalk on the fixed portion of Subaru Telescope Enclosure (44 m in diameter) (Fig.~\ref{fig:fig1}). The installation site's latitude and longitude are 19.82552, -155.47587, with a height of 13 m above ground. The camera is oriented approximately due east, encompassing the Keck I \& II telescopes, the NASA Infrared Telescope Facility (IRTF), the Canada–France–Hawaii Telescope (CFHT), the Gemini North telescope, and the University of Hawai'i 88-inch Telescope (UH88) within its field of view (Fig.~\ref{fig:fig2}). In this view, the horizon is situated about 5\% above the bottom edge of the frame, while structures within the summit area and mountain ridgelines constitute the majority of obstructions. Although unobstructed, a high-altitude view would simplify the processing of meteor observations, one of this camera's objectives is outreach, necessitating a field of view that includes the Maunakea Observatories. The obscuration rate of the field of view due to the horizon and structures is approximately 23\%. Table~\ref{tab:tbl1} summarizes information on the site location and the viewing area parameters.

\subsection{The Installation Site Condition}\label{ssec:2_3}
This section summarizes the observing conditions at the Maunakea summit, where StarCam is operated. Regarding Maunakea's high percentage of clear skies, reports indicate values exceeding 50\% for photometric nights and over 70\% for usable nights (allowing for some cloud cover), as detailed in \citet{1973PASP...85..255M} and the ESPAS Report (2003). Operational records from the Subaru Telescope also report an average of 18.3\% of the dome closure nights per year between 2021 and 2023\footnote{\url{https://www.naoj.org/Science/SubaruUM/index.html}}. Given that the Subaru Telescope often continues observations under conditions of low humidity even with significant cirrus clouds, this is consistent with the 70\% usable nights statistic. Therefore, it can be inferred that StarCam can be used for meteor observations for approximately 70\% of the nights.

The ambient temperature on the summit averages around 1.3 \celsius\, but can locally reach a maximum of 15 \celsius\ during the day, while dropping below 0 \celsius\ at night. Furthermore, temperatures can occasionally reach -10 \celsius\ during winter, making anti-icing measures for the window of the camera box crucial. Additionally, the significant temperature difference of the catwalk between day and night causes thermal expansion and contraction of the handrail where the camera is mounted, resulting in a slight shift in the field of view. However, the temperature is fairly stable at night, and no such shift is observed in the astrometric calibration data.

Direct sunlight, extreme dryness, and stronger ultraviolet radiation compared to lower altitudes accelerate the degradation of plastic components. It is necessary to select parts resistant to harsh environments. Additionally, elevated temperatures within the box, especially during summer, have previously led to protective camera shutdowns due to overheating. To mitigate this, an internal fan has been installed to automatically draw in external air when the internal temperature exceeds 20 \celsius.

\begin{table}
  \tbl{Camera Location.\footnotemark[$*$] }{%
  \begin{tabular}{cc}
      \hline
      Latitude & $19.8256~\degree$ \\ 
      Longtitude &  $-155.4758~\degree$\\ 
      Altitude of Camera & $4153~m$ \\
      Field of View (FoV) & $70~\degree \times 40~\degree$ \\
      FoV Center (Az \& El) & +81.2~\degree, 19.6~\degree  \\
      Sky Fraction & 77\% \\
    \hline
    \end{tabular}}\label{tab:tbl1}
\begin{tabnote}
\footnotemark[$*$] WGS84  \\ 
\end{tabnote}
\end{table}

\section{Details of the Camera System}\label{sec:3}
Here we describe the camera system in detail. Since its initial installation in 2021, the hardware has undergone three major upgrades along with continuous minor improvements. The specifications described below reflect the system's status as of 2025. We note that the major part of the system development was carried out by one of our Co-Is (M.H.). We briefly describe the history of the system update at the end of this section.

\subsection{StarCam System Design}\label{ssec:3_1}
\begin{figure}
\centering
\includegraphics[width=1.1\linewidth]{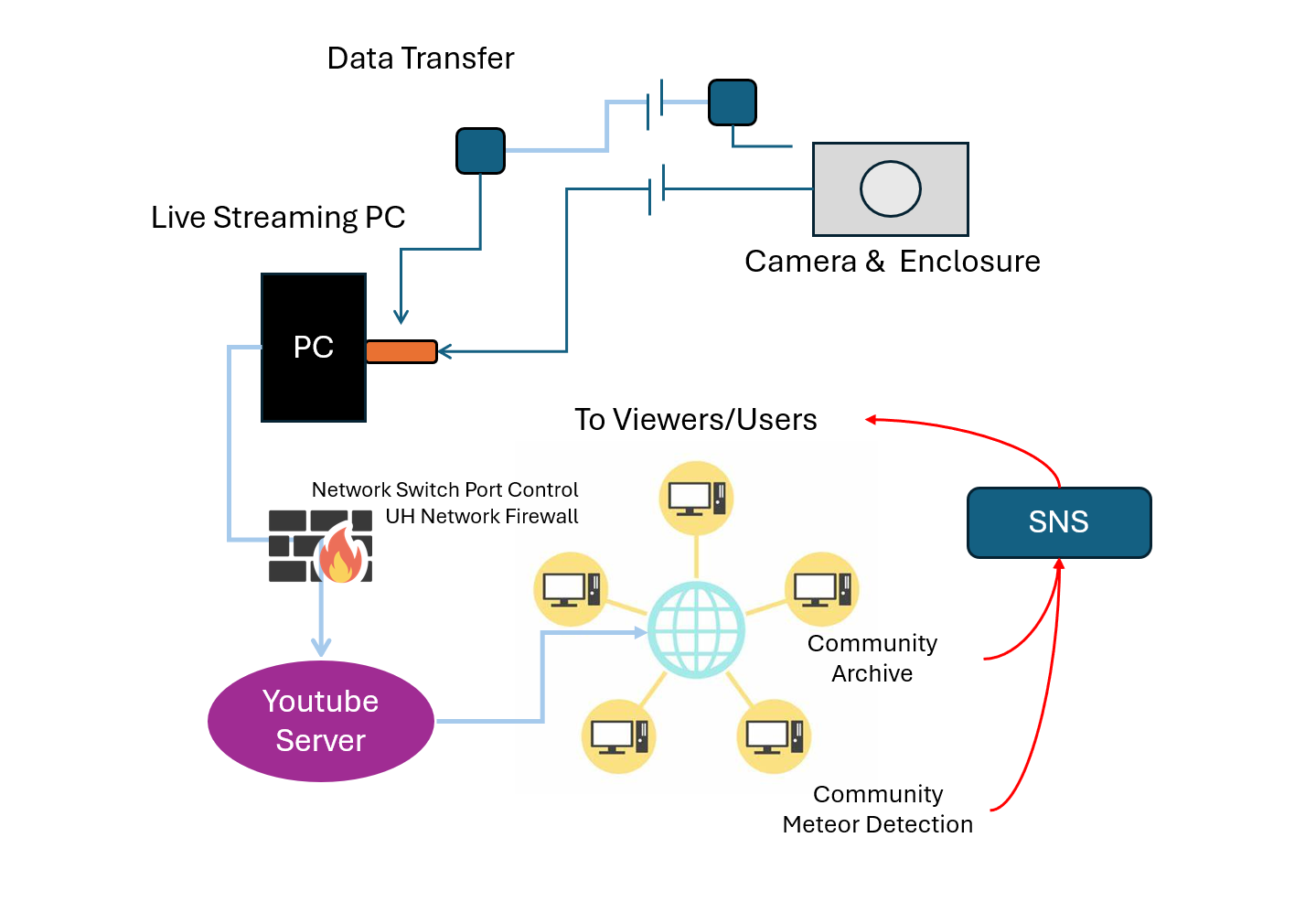}
\caption{Schematic diagram of the StarCam system. The video signal from the camera to the PC can be transmitted via two methods: HDBaseT Extender (top) or optical HDMI (bottom; current setup).
{Alt text: This figure is a schematic diagram of the StarCam system.} 
}\label{fig:fig3}
\end{figure}

Figure~\ref{fig:fig3} illustrates the system concept diagram. The camera, housed within an all-weather enclosure, is located over 30 meters away from the elevator tower where power and network connections are available. We have to send the video signal to the elevator tower. As it is a relatively long distance, the regular HDMI cables are ineffective. Moreover, measures against electrical noise are essential.

The received HDMI signal is captured by the PC there via a video capture device and subsequently sent to a YouTube server for live streaming. Viewers and data users are free to utilize the data for non-commercial purposes.

\subsection{The Camera}\label{ssec:3_2}
The camera and lens are central to our system. Our system employs a Sony FX3 (ILME-FX3) mirrorless cinema camcorder. The FX3 features a 10.2-megapixel full-frame backside-illuminated CMOS sensor with a native ISO range of 80 to 102400 with the optional sensitivity enhanced to ISO 409600. Table~\ref{tab:tbl2} summarizes the key setting parameters for the current live streaming (FX3).

The lens paired with the camera is a Sony FE 24mm F1.4 GM. This lens offers excellent corner-to-corner resolution, achieved through its optical design incorporating two extreme aspherical elements and ED (extra-low dispersion) glass. Given that the signals we capture are primarily point sources, the performance requirements for the lens are significant. The measured level of edge darkening at F1.4 is also acceptable.

A soft filter (Kenko PRO1D Clear Filter) is used with the lens to mitigate the saturation of bright objects, which occurs due to the 8-bit signal format used in live streaming. By employing the soft filter, we can gather more signal from brighter stars. This improves the recognition of constellations and simultaneously aids in the magnitude estimation of meteors.

\begin{table}
  \tbl{Key Camera Settings.\footnotemark[$*$] }{%
  \begin{tabular}{cc}
      \hline
      Exposure Mode & Program Auto \\ 
      ISO &  AUTO [range 80 to 409600]\\ 
      Exposure Comp. & +0.7EV \\
      Auto Slow Shutter & ON \\
      Focus Mode & Manual  \\
      Creative Look\footnotemark[$**$] & VV \\
      White Balance & C. Temp (4500K)\\
      HDMI Resolution & 2160p \\
    \hline
    \end{tabular}}\label{tab:tbl2}
\begin{tabnote}
\footnotemark[$*$] For the current camera (FX3).   \\ 
\footnotemark[$**$]  "Creative Look" is Sony's pre-defined color/tone setting for better visibility. The "VV" option is for an enhanced saturation and contrast for more colorful expression. \\ 
\end{tabnote}
\end{table}

\subsection{Exterior Box and Data Transfer}\label{ssec:3_3}
A unique aspect of the camera box's design is the inverted mounting of the camera, suspended from the ceiling. An aluminum plate is positioned at the top, and the camera is secured to this plate within the box via rigging. This design isolates the camera from the vibrations of the box caused by wind. Simultaneously, the aluminum top plate serves to protect the camera box from ice falling from the dome's ceiling, which is over 20 meters above. The flat-top design of the Subaru Telescope dome allows accumulated snow and ice to freeze into large masses. The impact of their fall can be substantial, making this safety measure critically important.

A 105 mm diameter camera lens protection filter is used for the box's window. An anti-reflective (AR) coating on the filter mitigates ghosting that could occur between the camera and the window. The filter can be rotated for attachment and detachment. While specific waterproofing measures are not implemented, Maunakea experiences low rainfall, which has fortunately prevented significant issues. A heater is attached to the window frame to prevent icing when the temperature drops below 0 \celsius.

The interior of the camera box frequently reaches considerably high temperatures. This is attributed to both the significant heat generated by the camera itself during 4K streaming, and the limited heat dissipation of the box due to direct sunlight and the reduced convective heat transfer caused by Maunakea's low atmospheric pressure. Because high temperatures can trigger the camera's built-in protection mechanism, leading to shutdowns, countermeasures are necessary. To address this, a ventilation fan was installed at the rear of the camera box to draw in external air when the temperature exceeds 20 degrees Celsius. Furthermore, the camera's thermal resistance has dramatically improved with the adoption of the FX3, which incorporates an internal fan mechanism.

However, the installation of the external air fan raises concerns about the potential for rainwater intrusion. Currently, measures such as installing a rain shield structure on the fan's exterior have not been implemented, but significant problems have not arisen. This is likely due to the camera box's rear facing the telescope dome wall, which shields it from direct wind exposure, and the fact that the annual precipitation is only about 15 cm, with the majority falling as snow. As a precautionary measure, drainage holes have been drilled into the bottom of the box's door to allow water to escape.

These modifications are a result of operational know-how gained from the Phase 1 and 2 periods (see sect.~\ref{ssec:3_5}). The initial camera system experienced frequent streaming disruptions due to thermal runaway and persistent issues with condensation. In contrast, these problems have been largely resolved in the current camera box.

The distance from the camera box to the nearest Ethernet port exceeds 30 meters. Due to the prohibition of Wi-Fi usage at the summit, the 4K 2160p \& 30 fps video signal output by the camera must be transmitted via a wired connection over a distance exceeding 30 meters. However, because of the long transmission distance, standard HDMI cables cannot transmit the signal effectively due to high-frequency digital signal attenuation.

To circumvent this, we implemented a long optical HDMI cable. Optical HDMI cables convert electrical HDMI signals into optical signals for transmission, which experiences extremely low loss over long distances. They are also immune to noise and crosstalk. We note that we previously utilized an HDMI Extender over Ethernet Cable (HDBaseT Extender). Both solutions functioned effectively, but optical HDMI cables offer much less trouble and offer greater simplicity as a system.

The power consumption on the camera box side is approximately 20 W for the camera and 10 W each for the fan and heater. The power source for the camera box is the commercial power supply in a building about 30 meters away. As the camera occasionally requires a power cycle to recover from an overheat shutdown, we are planning to add a network Power Distribution Unit for remote power cycling of the camera. We note that the necessity for such camera power cycles is at most a few times per year.

\subsection{The PC for Live Streaming}\label{ssec:3_4}
The video signal received by the PC via the optical HDMI cable is captured using a video capture card. The PC we utilize for live streaming is equipped with an Intel Core i5-13400F CPU and 16 GB of RAM. It also features a GeForce GTX 1660 Super GPU, providing ample specifications even for 4K streaming.

We utilized Streamlabs\footnote{\url{https://streamlabs.com/}} as our streaming software. Previously, we had been using OBS Studio\footnote{\url{https://obsproject.com/}}, but we switched due to compatibility issues with our capture device when we upgraded the PC. For 4K live streaming, we set a bitrate of 20,000 Kbps with a 1-second keyframe interval using H.264 hardware encoding.

Incorporating a timestamp into the viewing window of the live stream is crucial when we use the data for science, especially in meteor science. The PC's clock was designed to automatically synchronize every 12 hours via NTP. However, for unknown reasons, the NTP synchronization frequently fails to operate, resulting in unreliable timekeeping. Consequently, past data may exhibit time discrepancies exceeding 10 seconds.  Only recently has the automatic adjustment started working properly by editing the registry information.

\subsection{Update History}\label{ssec:3_5}
Since the start of live streaming on April 3, 2021, the StarCam system has undergone continuous updates, including two major camera replacements in the past. Therefore, we have summarized these periods, referring to them as Phase I, II, and III, in Table~\ref{tab:tbl3}.

Phase I had aspects of a test run, trying to find any issues related to streaming or summit camera operation. After a few months from the start of streaming, Phase II involved improvements to the camera box and an upgrade of the camera itself. A noteworthy event during Phase II was the introduction of a soft filter starting on February 18, 2022. Additionally, a short-term test of 4K streaming was conducted during this phase. Phase III is characterized by operation with the goals of further stabilization through hardware upgrades and the long-awaited realization of 4K streaming. The number of troubles has been significantly reduced by the introduction of current hardware.

\begin{table}
  \tbl{Update History of the StarCam.}{%
  \begin{tabular}{lcl}
  Phase I &2021-4-3 to 2021-8-8 & \\
      \hline
      Camera & Sony Alpha 7S II (APS-C) & SIGMA 16mm F1.4\\ 
      Streaming & HD (1080p,30fps) &  ISO 50 -- 204800 \\ 
      & & \\
  Phase II & 2021-8-9 to 2023-8-9 & \\
      \hline
      Camera & Sony Alpha 7S III & Sony 24mm F1.4 GM\\ 
      Streaming & HD (1080p,30fps)  &  ISO 80 to 204800 \\ 
      & & \\
  Phase III & 2023-8-10$\sim$Current  &\\
      \hline
      Camera & Sony FX3 & Sony 24mm F1.4 GM\\ 
      Streaming & 4K (2160p,30fps) &  ISO 80 -- 409600\\ 
    \hline
    \end{tabular}}\label{tab:tbl3}

\end{table}

\section{System Performance}\label{sec:4}
\subsection{Faint Object Detection Performance}\label{ssec:4_1}
\begin{figure}
\centering
\includegraphics[width=0.8\linewidth]{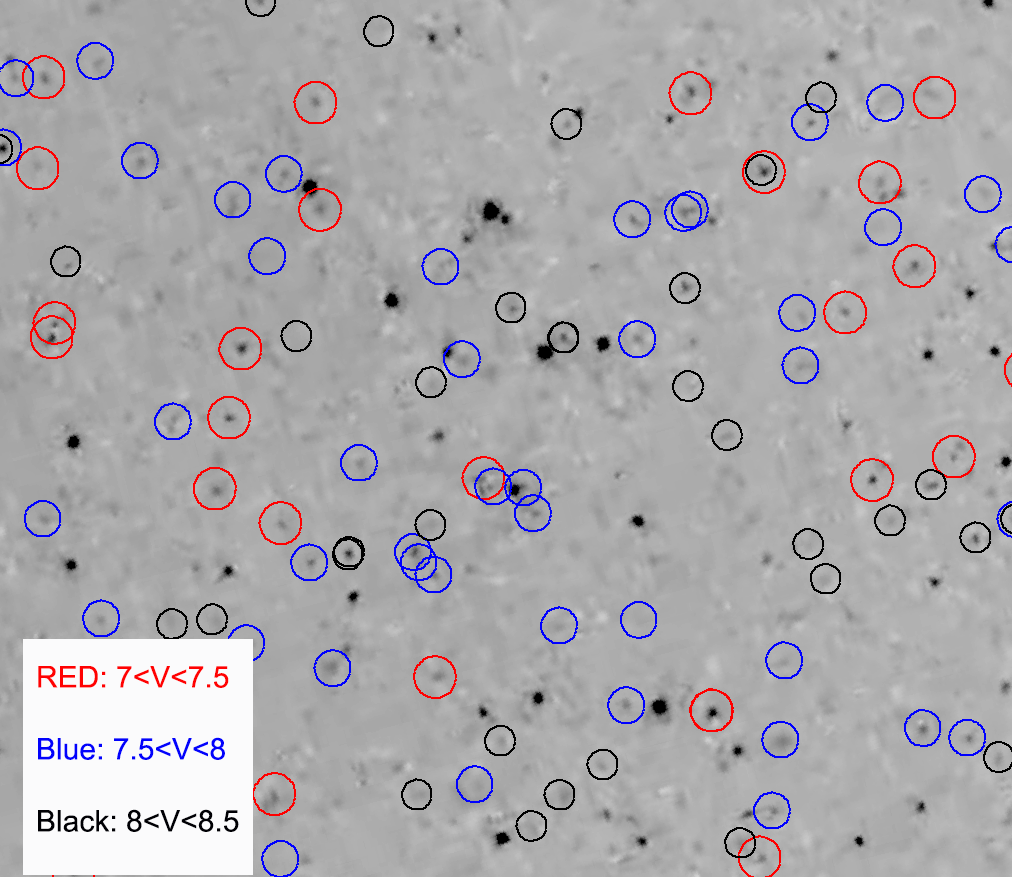}
\caption{Sample image from StarCam streaming data: a cropped single frame with an exposure time of 1/15 sec at ISO 409600, shown in inverted grayscale. Overlaid markers indicate the positions of faint stars (constellation Delphinus centered).
{Alt text: This figure demonstrates the faint end performance of the StarCam. } 
}\label{fig:fig4}
\end{figure}

To evaluate StarCam's faint-object detection capabilities, we analyzed a sample from a dark, clear night of streaming data recorded shortly after 21:00 HST on April 24, 2025. Under these dark-sky conditions, the camera operated at ISO 409600 with a shutter speed of 1/15 second, maintaining 30 fps. Figure~\ref{fig:fig4} shows an inverted grayscale image of a single frame from this data, with an overlay of stars from the Hipparcos Main Catalog ranging in V magnitude from 7 to 8.5. Almost all stars brighter than magnitude 8 are clearly visible, and a significant number of stars fainter than 8 mag are also detected.

\subsection{Vignetting}\label{ssec:4_2}
While the camera lens exhibits good performance, notable peripheral vignetting is observed due to its use at maximum aperture at night. Slight vignetting caused by the window's edge also occurred during the field of view adjustment upon installation. Ideally, flat frames would be used to assess peripheral vignetting. However, this is difficult due to the camera's permanent outdoor installation. Therefore, we utilized streaming data from a "whiteout" condition caused by dense fog at Maunakea (18:00 HST on January 30, 2025) as a substitute for flat frames in our evaluation.

\begin{figure}
\centering
\includegraphics[width=0.8\linewidth]{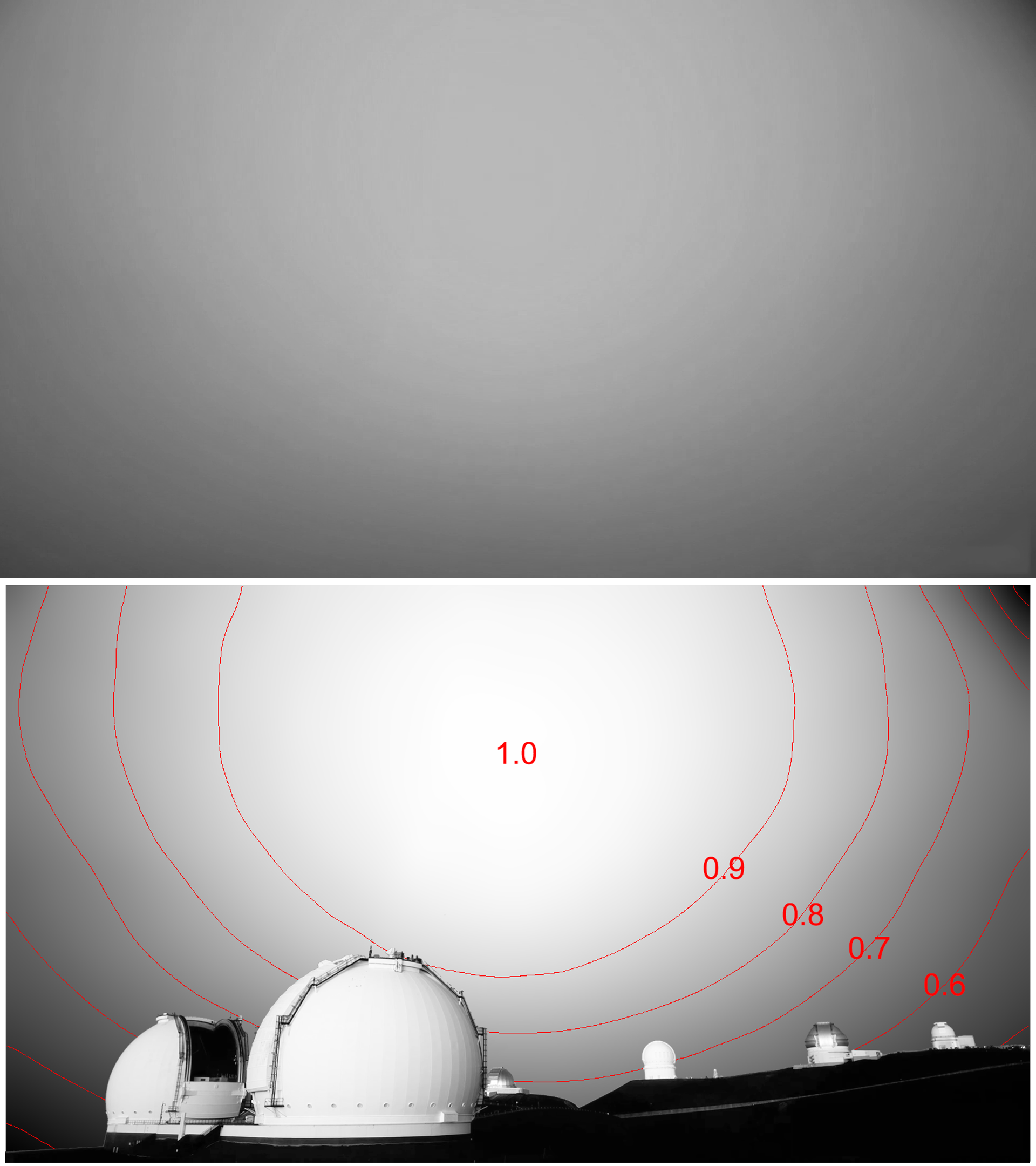}
\caption{Normalized “flat-field” image illustrating peripheral vignetting. The upper panel is the raw view, while the lower panel is with contours overlaid. The ground-view is also overlaid for the positional reference. Though the fog is so dense that it homogenizes the incoming light distribution, some effect from dark ground may remain. We will try taking a better flat images when we have a chance to remove camera box.
{Alt text: This map shows the degree of vignetting by the lens, the camera window, and so on. } 
}\label{fig:fig5}
\end{figure}

Figure~\ref{fig:fig5} shows the normalized brightness distribution. Note that while a clear-sky ground view is overlaid for positional reference, the actual data is uniformly white with no discernible features. However, we cannot rule out that the intrinsic darkness of the ground direction could still affect the homogeneity of the light source, especially in the sky region near the summit ridge. Based on this analysis, a flux loss of 30 to 40\% (corresponding to 0.4 to 0.5 magnitudes) could be happen near the edge of the field of view.

\subsection{Photometric Accuracy}\label{ssec:4_3}
To obtain an estimate of meteor brightness, we assessed the photometric accuracy from the streaming data. The primary challenges are the limited 8-bit dynamic range inherent in the MP4 format and the saturation of many bright stars. Significant photometric errors are likely unavoidable under these conditions.

For photometric calibration, we first divided a one-second video into individual frames and combined them using the median value of each pixel. Astrometry was then performed on this resulting image using astronomy.net\footnote{https://astrometry.net/}. To detect stars and perform photometric measurements, we used DaoStarFinder\footnote{\url{https://photutils.readthedocs.io/en/stable/api/photutils.detection.DAOStarFinder.html}}. We calibrated our photometry using V magnitudes from the Hipparcos Main Catalog\footnote{\url{http://vizier.u-strasbg.fr/cgi-bin/VizieR?-source=I/239/hip\_main}}. Stars were identified via positional cross-matching, enabling us to evaluate the offset between catalog and instrumental magnitudes. The instrumental magnitudes were corrected for vignetting, as assessed in the previous section.

\begin{figure}
\centering
\includegraphics[width=0.95\linewidth]{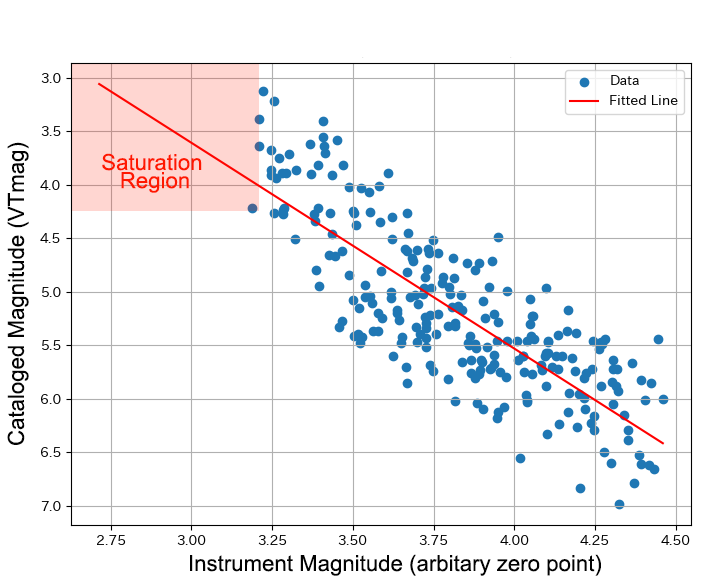}
\caption{Relationship between instrumental magnitudes (horizontal axis) and catalog magnitudes (vertical axis). Objects brighter than magnitude 4 are saturated in the StarCam video and thus omitted. See the text for more details.
{Alt text: The figure showing the relation between the instrument magnitude and the catalog mag. } 
}\label{fig:fig6}
\end{figure}

The results are shown in Figure~\ref{fig:fig6}, where the horizontal axis represents the instrumental magnitudes measured from the image, and the vertical axis represents the catalog V magnitudes. Note that the brighter end is excluded by DAOStarFinder due to saturation.

We note that this relationship is dependent on the ISO sensitivity and shutter speed. Therefore, for actual meteor photometry, a linear fit of this count-magnitude relationship is determined using frames immediately before and after the meteor's appearance in the video. This linear fit is then used for the photometric calibration of the meteor in each individual frame.

.
\section{Meteor Detection, Database, and Archive}\label{sec:5}
Since the start of the live streaming with StarCam in 2021, we consistently have a dedicated real-time viewership far exceeding 100 individuals at any given time each night, who enjoy the night sky while monitoring phenomena such as meteors. The high percentage of clear skies and dark nighttime conditions at Maunakea have enabled continuous meteor observations with minimal cloud interference. As a result, StarCam has enabled scientifically significant observations, including unexpected meteor shower outbursts and meteor clusters. It is worth emphasizing that visual detections and information sharing by these enthusiastic viewers have been crucial to these discoveries.

From a citizen science perspective, it is particularly noteworthy that the community has begun actively utilizing this streaming data as a shared scientific resource. Two notable examples include the automatic detection and distribution of meteor data via a dedicated database, and the voluntary creation of an archive by viewers. In the following, we describe the meteor database in more detail.

\begin{figure*}
\centering
\includegraphics[width=0.75\textwidth]{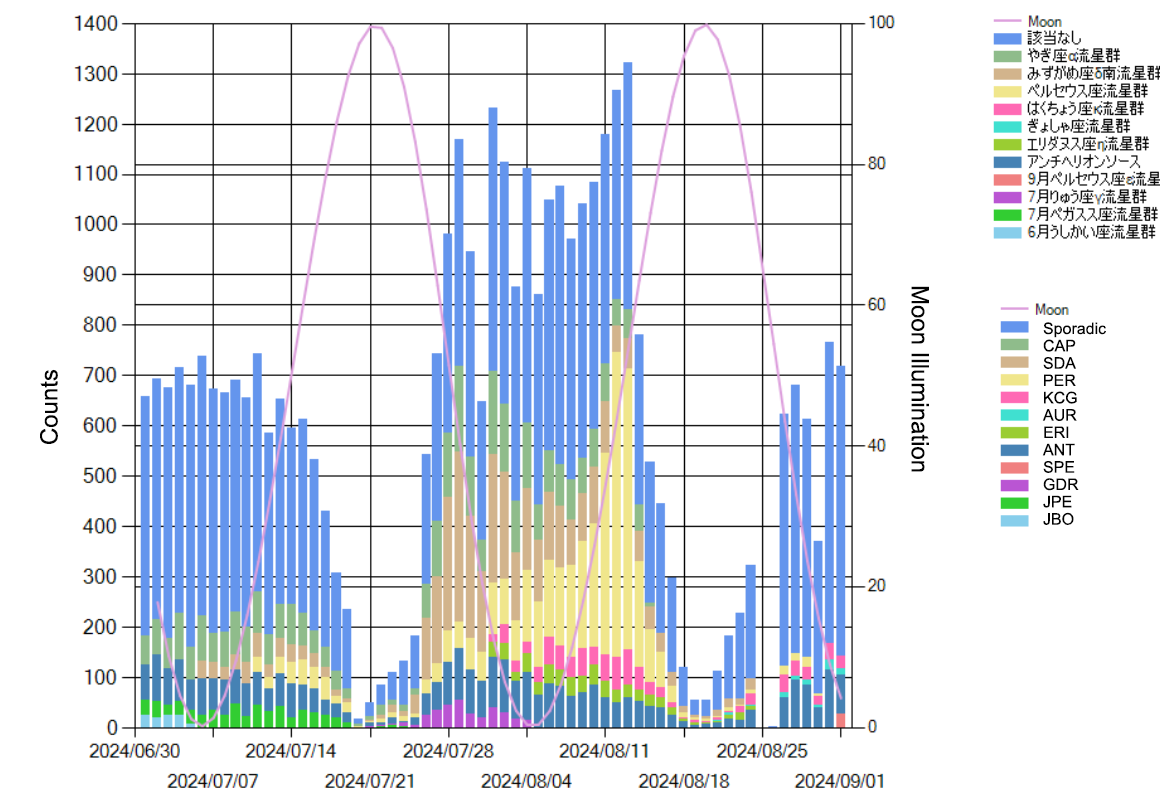}
\vspace{10mm} 
\includegraphics[width=0.75\textwidth]{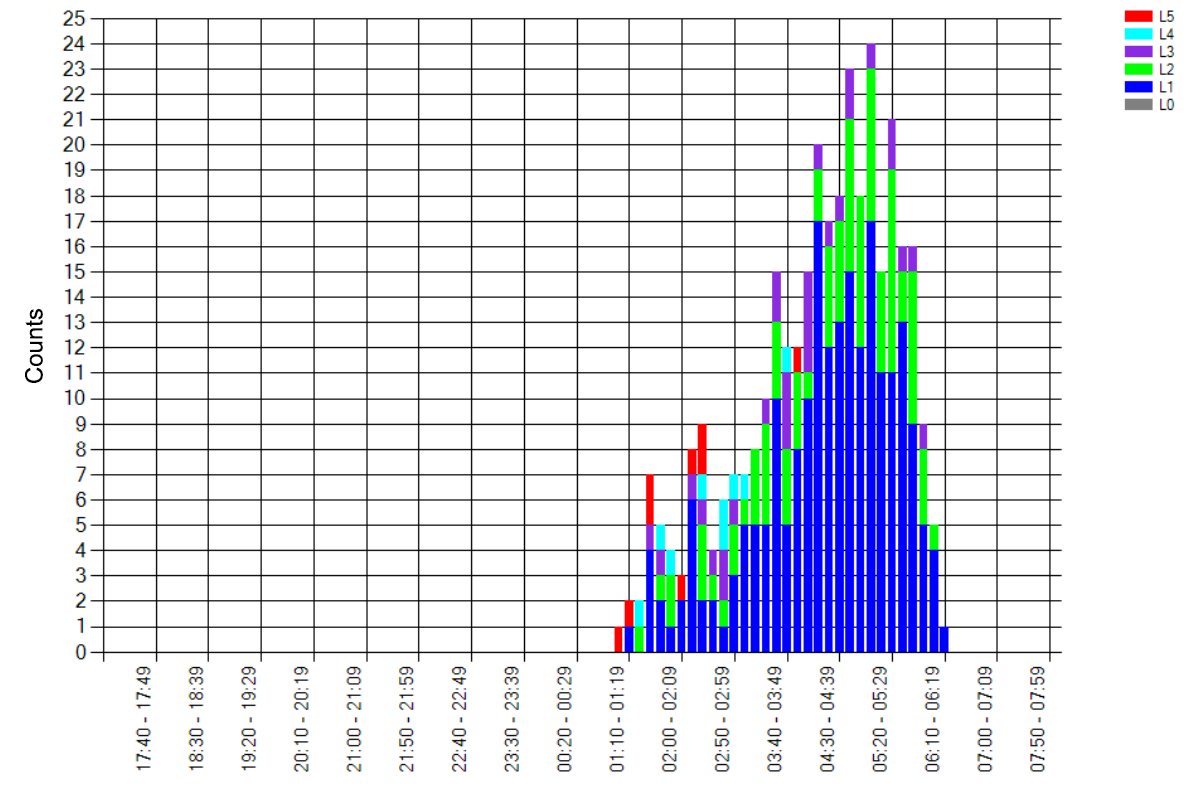}

\caption{[Top: An example of the daily histogram of meteors detected by StarCam, showing data from June 30th to September 1st, 2024. Meteors are color-coded by shower (Caption is in Japanese, so we put the IAU Shower Number here as an english translation of it below). The pink curve represents the lunar phase, illustrating a significant decrease in detected meteor counts during brighter moon phases. Prominent peaks are visible for the Delta Aquariids in late July and the Perseids on August 12th. Sporadic meteors are shown in light blue. 
Bottom: the histogram of the 10-min bin counts of  the Quadrantid meteor shower on January 2nd, 2025 HST. Color indicates meteor length, represented by L1 (shortest) to L5 (longest). The radiant rises in Hawaii at 1:00 AM HST. Earth-grazer-like meteors with long paths are evident when the radiant is low, with meteor frequency increasing as the radiant rises and ceasing at dawn.
{Alt text: The top figure shows the daily histogram of the detected meteors with date. Meteors are color-coded by shower. The bottom figure shows the histogram of the 10-min count of the Quadrantid meteor shower activity on January 2nd, 2025 HST.} 
}\label{fig:fig8}
\end{figure*}

A system that detects meteors from the YouTube live stream, performs characterization and astrometry, and publishes the results in an open database, has been developed and maintained by one of our co-investigators (TU). This "Hawaii Maunakea Meteor Database\footnote{\url{https://aiharasoft.com/stars}}" 
\ incluldes information of automatically-detected meteors such as X/Y, Right Ascension and Declination (J2000) for its start and end points, the meteor shower classification based on the trajectory analysis, approximate event time, average velocity, crude brightness estimate, color and so on. 
Additionally, screenshots of detected meteors are available for download, and videos of prominent events are shared via social media platforms. The system also allows for the graphical display of the time series of meteor counts, enabling rapid evaluation of meteor shower activity levels (Figure~\ref{fig:fig8} as a demonstration). The meteor detection system and database began operation in May 2022. As of April 2025, approximately 583,000 meteor entries have been recorded.

Alongside the database, the archive of the live video is also crucial for science. YouTube does not archive live streams longer than 12 hours. However, frequently starting and stopping the stream each night is impractical for outreach purposes, as the URL changes with each session. Therefore, it is necessary to record the stream view separately. The recording data for 4K, 30 fps video typically amounts to tens of gigabytes per hour, making daily archiving of the entire night sky impractical due to the enormous storage requirements.

\begin{figure}
\centering
\includegraphics[width=1.0\linewidth]{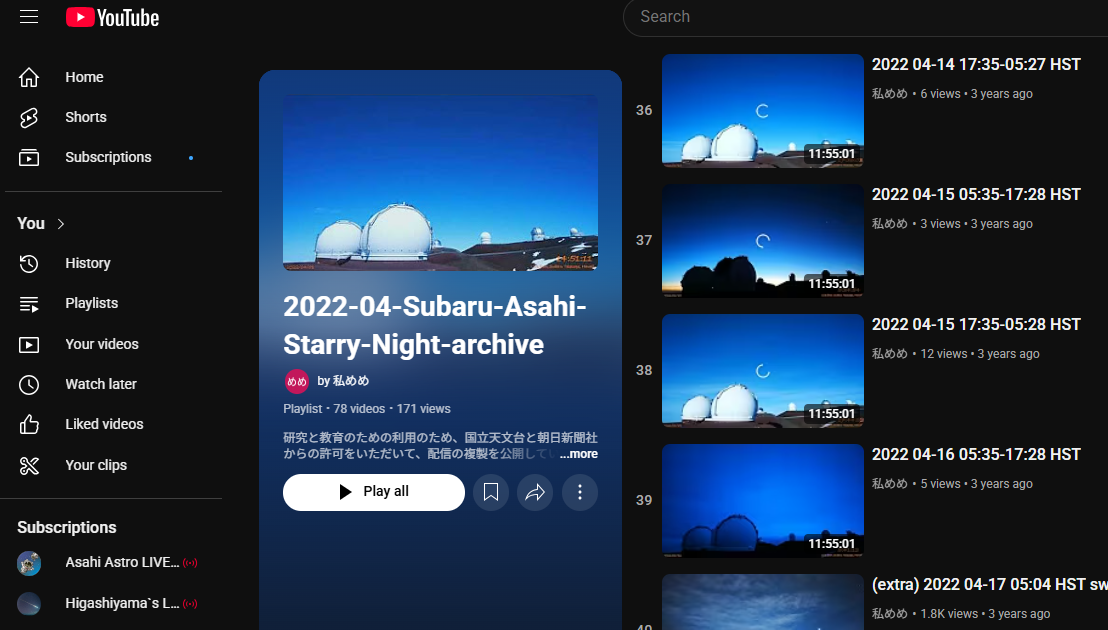}
\caption{Subaru-Asahi StarCam Video Archive, created and maintained by a community contributor, Watashi-Meme. This Youtube playlist can be found at \url{https://starcam.short.gy/meme_MKplaylists}. A detailed journal of the video recording data is also available (in Japanese). See Watashi-Meme (2023, 2025) for details.
{Alt text: This is a sample page of the StarCam video archive page. } 
}\label{fig:fig9}
\end{figure}

Under these circumstances, an engaged viewer voluntarily initiated an archiving project\footnote{\url{https://starcam.short.gy/meme_MKplaylists}}. They developed a system that records the live stream in HD (1080p) format every six hours and automatically uploads it to YouTube. This auto-archiving system has been in operation since September 2023. The recorded videos on Youtube are then reorganized into monthly playlists, allowing for quick access to revisit footage from a specific day (See Figure~\ref{fig:fig9}. This community archive began from September 2021 as 1--4 hr range video archives, and then expanded to longer-time recordings (6 to 12 hr) including daytime data from February 2022. All the details of the recording are maintained in separate documents for the user's convenience (\cite{watashimeme2023}, \cite{watashimeme2025}). This fully public, community-driven science infrastructure on the YouTube platform is highly distinctive.

\section{Science Highlightes}\label{sec:6}
Over four years of operation, StarCam has produced several notable scientific results in the field of meteor science. Some of these findings are discussed in detail in separate articles within this special issue.

One major result is the detection of meteor cluster phenomena. A meteor cluster refers to a group of meteors appearing almost simultaneously within a few seconds, first reported during the 2001 Leonid meteor shower (\cite{1999GeoRL..26...41K}, see also \cite{2003PASJ...55L..23W}). Due to the limited number of reported cases, meteor clusters are considered rare, and their frequency of appearance is yet to be established. Since the first detection of a meteor cluster by StarCam in July 2021, similar events have been observed nearly every year. This detection frequency is unprecedented in previous reports and is likely attributable to StarCam’s unique features: a high percentage of clear nights and a wide field of view that allows the detection of faint meteors. Watanabe et al. (2025, in this issue) will discuss this in detail.

In October 2021, following a prediction of a potential new meteor shower--the Arid Meteor Shower--we conducted a targeted observation campaign using StarCam and successfully detected the event. This is discussed separately in Tanaka et al. in this issue.

The Perseid meteor shower in August 2021 benefited from favorable conditions, with the new moon coinciding with its activity peak. An unexpected strong secondary peak was observed two days after the main Perseid peak. This sub-peak was not predicted, and its cause remains unknown, making this observation one of the crucial datasets for estimating the activity scale of the meteor shower.

Notably, a Japanese elementary school student who continuously observed this shower throughout August using StarCam successfully recorded this sub-peak activity, which attracted media attention in Japan\footnote{\url{https://subarutelescope.org/en/news/topics/2021/08/26/2983.html}}. The origin of this sub-peak remains unclear.

In November 2021, StarCam obtained data confirming activity from the Andromedid meteor shower. This meteor shower was active until the 19th century but is considered to have largely disappeared in the 20th century. A sudden outburst of activity was reported globally in 2021, and StarCam captured the activity during the declining phase of this event. This will be discussed in Fujiwara et al. (2025, in this issue).

\begin{figure*}
\centering
\includegraphics[width=0.9\linewidth]{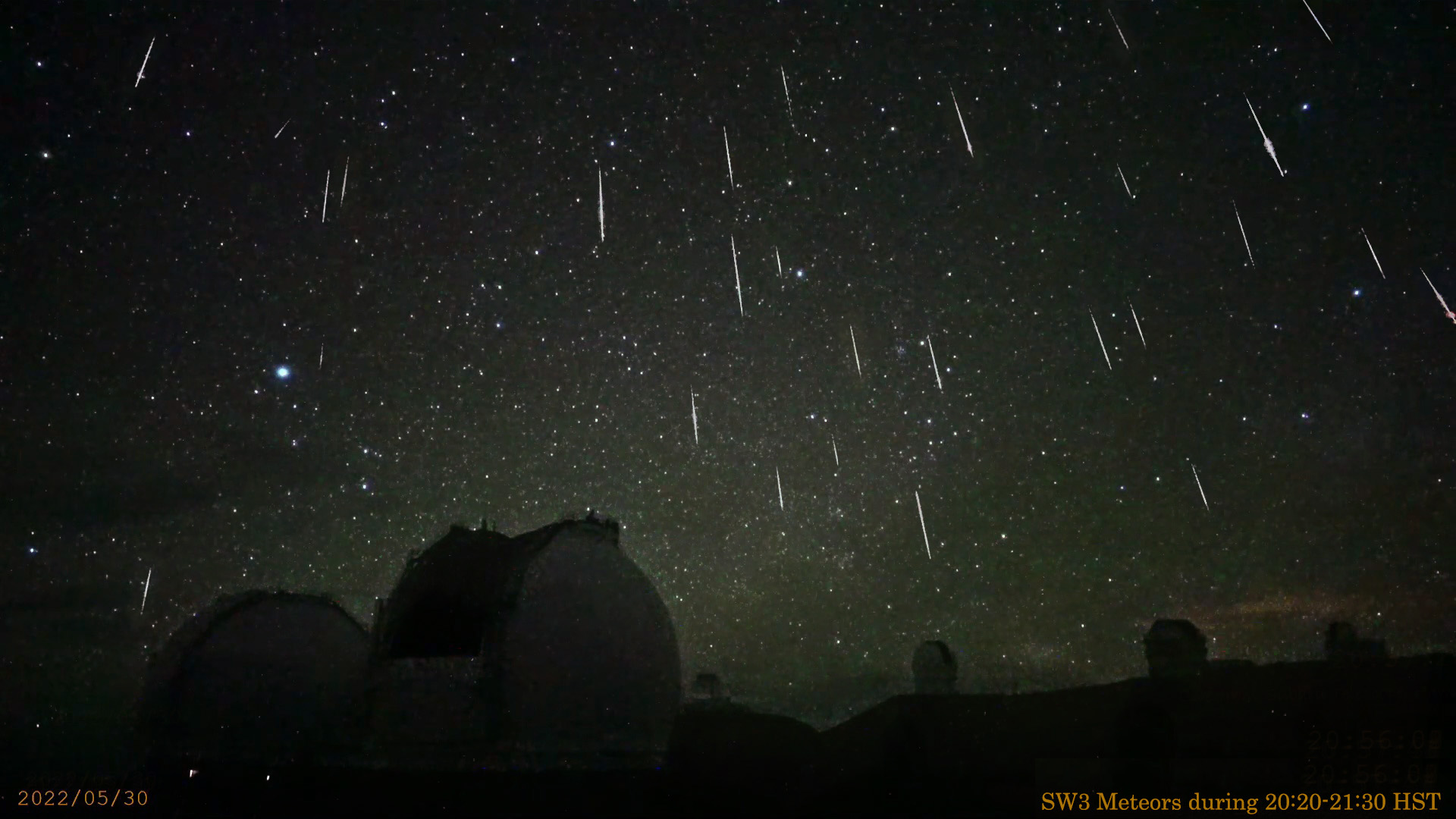}
\caption{A 70-minute composite image of the Tau Herculid meteor shower observed by the Subaru-Asahi StarCam (between 06:20 -- 07:30 UT, 31 May 2022). The image combines detected Tau Herculid meteors registered in the Maunakea Meteor Database (Credit: NAOJ \& Asahi Shimbun).
{Alt text: This is a stacked view of the Tau Herculids shower by Subaru-Asahi StarCam.} 
}\label{fig:fig10}
\end{figure*}

On May 31 2022 (UT), strong activity of the Tau Herculids meteor shower was predicted by our co-investigator (M.S.), and StarCam successfully recorded numerous red, slow-moving meteors after sunset (Fig.~\ref{fig:fig10}). The analysis results are discussed in Sato et al. (2025, in this issue).

The detection of a group of meteors having the same orbit as asteroid 2024 RW1, which entered the Earth's atmosphere and burned up in September 2024, has been reported by \citet{2024emetnj_sekiguchi}. Within this report, it was noted that the StarCam also captured multiple meteors believed to be of the same origin.

Beyond these achievements, there are several other noteworthy phenomena observed. The StarCam occasionally captures "parallel meteors," where two or three meteors simultaneously traverse the field of view in the same direction. These are also rare events, and their occurrence frequency and luminosity distribution remain unknown. Furthermore, the analysis of the annual trend on the occurrence of faint meteors (magnitude 5-6), which are likely difficult to detect with the naked eye, is a relatively new field, and StarCam data in this regard remains unanalyzed. Additionally, viewers have frequently noted a phenomenon known as "flickering meteors", in which a meteor dims and then brightens again before vanishing. The cause of such peculiar light curve variations, their occurrence frequency, and why we predominantly observe meteors dimming only once, with a complete absence of meteors dimming two or three times, representing potential research topics where StarCam could provide valuable insights.

\section{Discussion and Future Scope}\label{sec:7}
This paper has presented the hardware specifications of StarCam, installed on the Subaru Telescope at Maunakea, Hawai'i. The preceding section briefly summarized this camera's contributions to meteor science. It is important to emphasize that these achievements result not only from the advanced capabilities of the camera-—its wide field of view and high sensitivity—-but also from Maunakea's world-class observing conditions and the continuity of our long-term observations. Furthermore, Hawai'i's unique geographic location in the mid-Pacific provides significant scientific advantages as an observing site. Hawai'i offers a valuable and irreplaceable vantage point for understanding the time series of meteor shower activity, one that cannot be readily substituted by other observing sites.

Furthermore, as previously discussed, our methodology to publicly stream this data on YouTube for outreach purposes has unexpectedly fostered citizen science driven by a large and engaged viewership. The ability to share the excitement of meteor research with viewers of the live sky feed is itself a major achievement of the StarCam initiative.

On the other hand, there are limitations inherent in this camera's primary objectives: monitoring the meteorological environment of the astronomical observing site and outreach through sharing the night sky of Maunakea.

If dedicated meteor observation were the primary goal, a monochrome camera with high linearity would be preferable, and the frame should exclude structures and the horizon. Pointing towards the zenith would also simplify the conversion of counts to ZHR. However, this approach would not serve the outreach purpose of allowing many people to be moved by the natural beauty of Maunakea captured in the imagery, or to be impressed by the sight of the night sky and meteors.

Conversely, this StarCam serves as a good demonstration of the power that comes from placing a flagship camera at Maunakea, a world-class observing site.

Maunakea is a sacred site. Unfortunately, the installation of new equipment dedicated solely to astronomical observation, even a small camera lens, is currently not possible. This is due to the stringent restrictions that limit astronomical telescopes on Maunakea. Even a small camera lens, if the system is dedicated to astronomy, needs to be passed through a lengthy permission process. Therefore, it is important to shift our perspective and consider how scientific output can be derived from a camera primarily intended for other purposes, like outreach or site monitoring.

Importantly, the fact that we publicly share first-class starry sky and meteor observation data on open platforms like YouTube has led to unexpected citizen science discoveries through public scrutiny of the live view. StarCam demonstrates that even familiar phenomena like meteors can yield novel discoveries when examined through accessible data such as live streams. A wealth of information regarding meteors still remains untapped within our database. We hope that by effectively utilizing these resources, the data from this camera can make further significant contributions to meteor science.

We would like to discuss the future of StarCam. We recently installed and began operating a StarCam at CFHT (Jan 2025). This camera monitors the western sky. Furthermore, preceding this in 2022, the PANOPTES Project (\cite{2022ASPC..533..217K}), based on Mauna Loa Observatory, also implemented a StarCam system. Unfortunately, shortly after we began test operation, the eruption of Maunaloa occurred in November 2022, destroying the access road and the power and network infrastructure to the camera site. Regrettably, the site has not yet been restored, but we plan to resume operation of this Maunaloa StarCam once power and network connectivity are re-established.

Once the Maunakea StarCam and the Mauna Loa StarCam can be operated collaboratively, it will become possible to calculate the orbital information of meteors through double-station observations. The camera sites have a north-south separation of 34 km. Considering the typical meteoroid ablation height ($\sim100$ km) and the observed sky area of each StarCam, this distance is appropriate for determining position and velocity via triangulation. Determining meteoroid orbits is fundamental to meteor science, and this capability is expected to greatly enhance the scientific value of StarCam data. The dissemination of such open science resources from Hawai'i can also contribute to the inspiration of citizen science within the state. Broad outreach to local educators regarding our activities will be an important next phase.

Situated in the middle of the Pacific Ocean, Hawai'i is a crucial site for meteor observation networks. The state's span of over 500 km also provides an appropriate scale for developing a center for meteor science based on the network of multiple observing stations across the islands. Starting with StarCam as a first step, our aspiration is to eventually construct a larger Hawaiian meteor network encompassing other islands as well.

\begin{ack}
First and foremost, we thank our committed viewers of Astro LIVE by the StarCams. Their invaluable contributions have been essential to many of our discoveries. Furthermore, we extend our utmost gratitude to `Watashi-Meme' for her long dedication to the development of the voluntary archive. This contribution supports one of the scientific foundations of StarCam and is indeed worthy of co-authorship. We are deeply grateful to the Subaru Telescope Directorate, especially former Director Dr. Michitoshi Yoshida and current Director Dr. Satoshi Miyazaki, for their generous support in initiating and maintaining the Subaru live stream. We also appreciate the Center for Maunakea Stewardship for permitting the operation of this camera on the summit. We thank Mr. Naotsugu Mikami for sharing his expertise in both the software and hardware aspects of live streaming. We also express our sincere thanks to Mr. Chris Chock and Dr. Olivier Guyon for their efforts in optimizing 4K live-streaming parameters for the Maunaloa StarCam. We gratefully acknowledge the Subaru Computer Division, led by Mr. Kiaina Schubert, for their professional support, as well as the Subaru day crew and operations group for their assistance with on-site troubleshooting. Finally, We thank Dr. Luk\'a\v{s} Shrben\'y, the referee of our paper, for his insightful comments and support for publication.  The authors respectfully recognize the profound cultural significance and enduring reverence that the summit of Maunakea holds for the indigenous Hawaiian community. We feel deeply honored to have the opportunity to conduct astronomical observations from this sacred mountain.

\end{ack}



\begin{thebibliography}{}
\bibitem[Borovi{\v{c}}ka et al.(2021)]{2021M&PS...56..425B} Borovi{\v{c}}ka, J., Bettonvil, F., Baumgarten, G., et al.\ 2021, Meteoritics \& Planetary Science, 56, 3, 425.

\bibitem[ESPAS (2003)]{ESPAS_MK}
ESPAS site summary series: Mauna Kea (2003) http://www.eso.org/gen-fac/pubs/astclim/espas/espas\_reports/ESPAS-MaunaKea.pdf

\bibitem[Fujiwara et al.(2025)]{fujiwara2025sp} Fujiwara, Y., Tanaka, I., Hasegawa, H., \& Sato, M.\ 2025, \pasj, to be submitted soon (this issue)

\bibitem[Gomez \& Fitzgerald(2017)]{2017AstRv..13...28G} Gomez, E.~L. \& Fitzgerald, M.~T.\ 2017, The Astronomical Review, 13, 1, 28.

\bibitem[Jenniskens et al.(2011)]{2011Icar..216...40J} Jenniskens, P., Gural, P.~S., Dynneson, L., et al.\ 2011, \icarus, 216, 1, 40.

\bibitem[Jenniskens et al.(2012)]{2012Sci...338.1583J} Jenniskens, P., Fries, M.~D., Yin, Q.-Z., et al.\ 2012, Science, 338, 6114, 1583.

\bibitem[Kinoshita et al.(1999)]{1999GeoRL..26...41K} Kinoshita, M., Maruyama, T., \& Sagayama, T.\ 1999, \grl, 26, 1, 41.

\bibitem[Krishnamoorthy et al.(2022)]{2022ASPC..533..217K} Krishnamoorthy, P., Walawender, J., Gee, W.~T., et al.\ 2022, ASP 2021: Sharing Best Practices - AstronomyTeaching and Public Engagement, ASP Conference Series, vol 533, 217. 

\bibitem[Morrison et al.(1973)]{1973PASP...85..255M} Morrison, D., Murphy, R.~E., Cruikshank, D.~P., et al.\ 1973, \pasp, 85, 505, 255.

\bibitem[Sato et al.(2025)]{sato2025sp} Sato, M., Watanabe, J., Tsuchiya, C., Hasuo, R., Hasegawa, H., Tanaka, I., Uda, T., et al.\ 2025, \pasj, submitted (this issue). 

\bibitem[Sekiguchi et al.(2024)]{2024emetnj_sekiguchi} Sekiguchi, T. (2024), eMetN Meteor Journal, November, p. 403 

\bibitem[SonotaCo(2009)]{2009JIMO...37...55S} SonotaCo\ 2009, Journal of the International Meteor Organization, 37, 2, 55. 

\bibitem[Tanaka et al.(2025)]{itanaka2025arid} Tanaka, I., Sato, M., Hasegawa, H., Watanabe, J.-I. \& Higashiyama, M.\ 2025, \pasj, to be submitted soon (this issue)

\bibitem[Trigo-Rodr\'{i}guez et al.(2010)]{TR2010} Trigo-Rodr\'{i}guez, J.~P., Llorca, J., Madiedo, J.~M., Tancredi, G., Edwards, W.~N., Rubin, A.~E. \& Weber, P.\ 2010, Meteoritics \& Planetary Science, 45, 383.

\bibitem[Vida et al.(2021)]{2021MNRAS.506.5046V} Vida, D., {\v{S}}egon, D., Gural, P.~S., et al.\ 2021, \mnras, 506, 4, 5046.

\bibitem[Watanabe et al.(2003)]{2003PASJ...55L..23W} Watanabe, J.-I., Tabe, I., Hasegawa, H., et al.\ 2003, \pasj, 55, L23.

\bibitem[Watanabe et al.(2025)]{watanabe2025sp} Watanabe, J.-I., Hasegawa, H., Tanaka, I., \& Sato, M.\ 2025, \pasj, submitted (this issue)

\bibitem[Watashi-Meme (2023)]{watashimeme2023} Watashi-Meme, Subaru-Asahi StarCam Recording Data Archive Summary (2021-2023), \url{https://starcam.short.gy/archive_old}

\bibitem[Watashi-Meme (2025)]{watashimeme2025} Watashi-Meme, Subaru-Asahi StarCam Recording Data Archive Summary (2023-current), \url{https://starcam.short.gy/archive_new}

\bibitem[Weryk et al.(2013)]{2013Icar..225..614W} Weryk, R.~J., Campbell-Brown, M.~D., Wiegert, P.~A., et al.\ 2013, \icarus, 225, 1, 614.

\bibitem[Zuluaga et al.(2013)]{2013arXiv1303.1796Z} Zuluaga, J.~I., Ferrin, I., \& Geens, S.\ 2013, arXiv:1303.1796.


\end{thebibliography}

\end{document}